\documentclass[12pt,preprint]{aastex}

\shorttitle{Thermal emission of CoRoT-1b and WASP-12b}
\shortauthors{Zhao et al.}

\begin{document}


\title{Ground-based detections of thermal emission from CoRoT-1b and WASP-12b}


\author{Ming Zhao\altaffilmark{1, 2}, 
John D. Monnier\altaffilmark{3},
Mark R. Swain\altaffilmark{1},
Travis Barman\altaffilmark{4},
Sasha Hinkley\altaffilmark{5, 6}}

\email{ming.zhao@jpl.nasa.gov}

\altaffiltext{1}{Jet Propulsion Lab, 4800 Oak Grove Dr., Pasadena, CA 91109}
\altaffiltext{2}{NASA Postdoctoral Fellow}
\altaffiltext{3}{Department of Astronomy, University of Michigan, Ann Arbor, MI 48104}
\altaffiltext{4}{Lowell Observatory, 1400 W. Mars Hill Road, Flagstaff, AZ 86001}
\altaffiltext{5}{Department of Astronomy, California Institute of Technology}
\altaffiltext{6}{Sagan Fellow}


\begin{abstract}

We report a new detection of the $H$-band thermal emission of CoRoT-1b  and two confirmation detections of the $Ks$-band thermal emission of WASP-12b at secondary eclipses.
The $H$-band measurement of CoRoT-1b shows an eclipse depth of 0.145\%$\pm$0.049\% with a 3-$\sigma$ percentile between 0.033\% -- 0.235\%. This depth is consistent with the previous conclusions that the planet has an isothermal region with inefficient heat transport from dayside to nightside, and has a dayside thermal inversion layer at high altitude. The two  $Ks$ band detections of WASP-12b show a joint eclipse depth of 0.299\%$\pm$0.065\%. This result agrees with the measurement of Croll \& collaborators, providing independent  confirmation of their measurement. 
The repeatability of the WASP-12b measurements also validates our data analysis method.
Our measurements, in addition to a number of previous results made with other telescopes, demonstrate that ground-based observations are becoming widely available for characterization of atmospheres of hot Jupiters.

\end{abstract}





\section{Introduction}

Detecting thermal emission from transiting planets at secondary eclipses is a powerful technique to study their  atmospheres without spatially resolving them from their host stars. Recently, ground-based observations have emerged as a powerful tool, in addition to space-based observations, to characterize the atmospheres of transiting hot Jupiters \citep[e.g.,][etc.]{Sing2009, Rogers2009, Lopez-morales2010, Croll2010a}. These ground-based photometry observations mostly provide measurements at the $z', J, H, \&~ K$ band, with the near-IR ($J, H, K$) measurements probing deeper and higher-pressure layers  of planetary atmospheres than the Spitzer measurements \citep{Burrows2008}. Since the bulk of energy from hot Jupiters emerges from the near-IR between 1-3 $\mu m$ \citep{Barman2008}, ground-based observations have the potential to provide an important probe of these atmospheres.

CoRoT-1b and WASP-12b are two  Very Hot Jupiters that are well suited for ground-based secondary eclipse detections. 
CoRoT-1b \citep{Barge2008}  is the first planet that has emergent flux detected in both optical \citep{Snellen2009, Alonso2009} and the near-IR \citep{Gillon2009, Rogers2009}.  Recently, its thermal emission at 3.5  and 4.6$\mu m$ has also been measured by the Spitzer telescope \citep{Deming2011}. These studies provided multi-wavelength constraints of CoRoT-1b's atmosphere, suggesting it has a temperature of $\sim2310$K and possesses a thermal inversion layer at high altitude, consistent with the pM-class classification of hot Jupiters \citep{Fortney2008}. CoRoT-1b has also been shown to have an inflated radius of 1.45$R_{jup}$ \citep{Gillon2009}, which could be caused by tidal heating.  
However, its orbital eccentricity has been shown to be consistent with zero \citep{Rogers2009, Deming2011}, thus not providing any support for the theory that CoRoT-1b's inflated radius is due to the effects of tidal heating.

Despite the extensive studies of CoRoT-1b,  current atmospheric models still under-predict its $Ks$ and 2.09$\mu m$  measurements \citep{Gillon2009, Rogers2009}, and a blackbody model provides the best fit to the measurements \citep{Deming2011}. 
Additional measurements at other bands such as $H$ are desirable to further constrain CoRoT-1b's models and help us better  understand its atmospheric characteristics. 



WASP-12b is one of the hottest and most inflated planets discovered to date \citep{Hebb2009}. Powerful irradiation from its host star heats the atmosphere temperature above $2500$K \citep{Madhu2011}, resulting in deep eclipse depths that are favorable for ground-based secondary eclipse detections. Currently, thermal emission of WASP-12b has been detected in the $z', J, H, K$ bands and the Spitzer 3.5$\mu m$ and 4.6$\mu m$ bands \citep{Lopez-morales2010, Croll2011, Campo2011}. Based on these  observations, recent models of \citet{Madhu2011} suggest that the atmosphere of WASP-12b is extremely carbon-rich (specifically the C/O ratio of WASP-12b is $>$1 at 3-$\sigma$ significance). However, it lacks a prominent thermal inversion layer at photospheric depths predicted for very-hot Jupiters and has very efficient day-night heat redistribution \citep{Fortney2008}. These characteristics motivate new planetary interior models and present challenges to theoretical classifications of hot-Jupiter atmospheres \citep{Madhu2011}. 

Here we report a new detection of CoRoT-1b's thermal emission at the $H$ band with the Palomar 200in telescope, and two confirmation detections of WASP-12b's $Ks$ band emission with the Michigan-Dartmouth-MIT (MDM) 2.4m telescope. 
In \S\ref{obs} we present our observations and data reduction procedure. We discuss our data analysis process and the results of CoRoT-1b and WASP-12b in \S\ref{result}. In \S\ref{discuss} we compare the eclipse depth of CoRoT-1b with existing models,  and finally we summarize our results in \S\ref{conclusion}.

\section{Observations and data reduction}
\label{obs}

\subsection{CoRoT-1}
The observation of CoRoT-1b was conducted in the $H$ band with the WIRC instrument \citep{Wilson2003} on  Palomar 200-in Hale telescope on UT 2011 January 29. 
The WIRC camera has a 2048 x 2048 Hawaii-II HgCdTe detector with 
a pixel scale of 0.2487$''$/pixel and a wide field of view (FOV) of $8.7' \times 8.7'$. 
The observation started about 10.5 minutes before the ingress and ended about 102.5 minutes after the egress, lasting for about 255 minutes in total. Each image was taken with 30sec exposure and 1-fowler sampling. A total of 376 images were obtained. The duty cycle of the observation is 73\%. 
The average seeing of the night was  $\sim0.8"$.
We stayed on the target for the whole period without dithering to minimize instrument systematics. The telescope was slightly defocused to a FWHM of $\sim1.5''$ - $2''$ to keep the counts well within the linearity regime and to mitigate any potential intra-pixel variations. 
We adjusted the telescope occasionally throughout the observation to maintain the defocus. Because WIRC has no dedicated guider, the telescope pointing drift was 
manually corrected during the observation. 

For the data reduction process, we first subtracted all images with corresponding averaged dark frames. 
Sky flats were normalized and averaged to get a master flat field. We create a bad pixel mask with the master flat and dark frames. 
 The bad pixels in each image  were interpolated with cubic splines based on adjacent  flat-fielded pixels. 
 After these steps, stars within the flux range of 0.25 to 1.5 times of that of CoRoT-1 are selected as references. Stars with higher fluxes are beyond the linearity regime of the detector and thus are excluded. Stars with lower fluxes have insufficient signal-to-noise and are excluded as well. In addition, 4 stars within this flux range are excluded due to their excessive flux fluctuations compared to other stars. The selection leads to 31 well separated and evenly distributed reference stars in the FOV for  flux calibration. Due to the correlation of stellar flux variations  with their centroid positions on the detector (see \S\ref{CoRoT}), more reference stars are preferred to fewer stars in order to average out their correlation with centroid positions on the detector. 
  We calculated the centroids of all stars in each image using  a center-of-mass calculation, since it provided the smallest scatters of their relative positions.  
 The time series of CoRoT-1's centroid was determined by averaging the relative positions of all reference stars after correcting for 
 their relative distances. The resulting 1-$\sigma$ precision of the centroid determination is $\sim0.3$ pixels.

\begin{figure}[t]
\begin{center}
 \includegraphics[width=3.2in, angle=90]{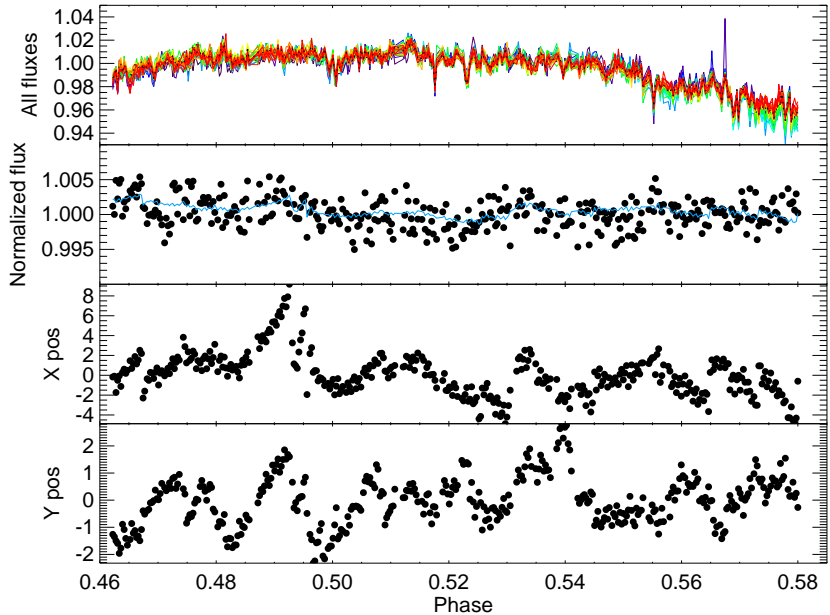}   
 \caption{Reduced fluxes and relative positions of CoRoT-1's centroid. The first panel shows the normalized fluxes of all 32 stars, including CoRoT-1 (black dotted line in the top panel). The second panel shows the final CoRoT-1 data normalized with the median of all reference fluxes. The two bottom panels show the relative positions of CoRoT-1's centroid. The centroid varies by $>$10 pixels during the observation. The blue line in the second panel shows the best-fit de-correlation model combined with a best-fit light curve (de-correlated with X positions only, see \S\ref{CoRoT}). The flux  variations in the top panel of Figure \ref{flux} are due to systematic effects such as airmass, seeing variations, and fluctuations of the atmospheric transmission, etc.   
 }
\label{flux}
\end{center}
\end{figure}

Aperture photometry was performed on CoRoT-1 and the reference stars following the IDL routines of DAOPHOT. 
The extracted fluxes of each star are normalized to the median of the time series. The median\footnote{Due to the presence of outliers in the light curves of some reference stars, as can be seen in Figure 1, we use median here as a more robust estimator.} of the 31 reference time series is then taken as the final reference light curve, which is used later to normalize the flux of CoRoT-1 to correct for the common-mode systematics such as variations of atmospheric transmission, change of seeing and airmass, etc.  We applied 48 different aperture sizes with a step of 0.5 pixels, and determined that an aperture with a radius of 9.5 pixels (19-pixel diameter) gives the smallest out-of-eclipse and in-eclipse scatters for the normalized CoRoT-1 data; this is  taken as the final photometry aperture for all stars in every image. Apertures within radii of $9.5\pm1.5$ pixels show consistent eclipse depths in later analysis, while apertures with larger than 1.5-pixel differences start to show excessive systematic noises. 
A sky annulus with 35-pixel inner radius and 30-pixel width was used for background estimation. The median value of the sky annulus was then used as the final sky background for subtraction. We have also explored different annulus ranges and sizes, and found consistent results.
The top two panels of  Figure \ref{flux} show the reduced fluxes of all 32 stars and the final normalized flux of CoRoT-1. 
The relative centroid changes of CoRoT-1 are shown in the two bottom panels. 
The UTC timestamp of each image was converted to BJD$_{UTC}$ first \citep{Eastman2010}, and then converted to orbital phases based on the ephemerides of \citep{Bean2009}, i.e., period=1.5089656 days, and transit epoch T$_0 (BJD_{UTC})$=2454159.452879. 

\subsection{WASP-12}

The observations of WASP-12b were conducted on two nights (UT 2010 November 26 \& 27) at the $Ks$ band with the TIFKAM  
imager on the MDM 2.4m Hiltner telescope. TIFKAM has a 1024x1024 HAWAII-1R HgCdTe detector.  
We used the f/7.5 imaging stop with a pixel scale of $0.2''$/pixel and a FOV of $3.4'\times3.4'$. The detector has numerous dead and hot pixels. We therefore carefully selected two ``clean" areas on the detector for both the target and the reference star. 
TIFKAM also has a known residual charge problem that each new image typically has a residual of 0.5-2\% of the previous signal. The residual can be reduced to $<<$1\% of the original signal when reading out the array several times with minimum exposure (see the TIFKAM manual). Therefore, for each science exposure in our observations, we readout the detector 3 more times with minimum exposure of 4.29sec and discard the 3 residual frames. The resulting duty cycles are roughly about 68\%, 58\% and 46\% for 90,  60, and 30 seconds exposures respectively.  

Our first observation started on UT 2010 November 26 at 05:49:55, about 19 minutes after the mid-eclipse, and ended at UT 09:10:41, about 134 minutes after the egress. 
A loss of telescope pointing occurred at UT 06:06:00, causing a gap of 17.63 minutes to the observation.  
We started with 60sec exposures and then reduced to 40sec to keep the flux within the linearity regime as the airmass decreased.
 We kept the telescope in focus and stayed on the target throughout the observation to minimize instrument systematics. A total of 118 images were recorded. The guiding precision was better than 3 pixels for both R.A. and Declination.  

The second observation started on UT 2010 November 27 at 06:17:50, about 2 minutes after the ingress, and ended on UT 11:38:04, about 149 minutes after the egress. 
The observations on both nights started after the beginning of ingress due to the late rise of WASP-12 and technical problems such as pointing \& guiding calibration, target centering, etc. We started with 90sec exposures  for the second night and then reduced to 30sec as airmass decreased. A total of 235 images were obtained. The telescope was also kept in focus. The guiding drift was about 3 pixels in Declination and 6 pixels in R.A..  The average seeing was $\sim1.5''$ for both nights. 

Reduction of the WASP-12 data followed the same procedures described above for CoRoT-1. 
Due to the sparse field of WASP-12, we only found one good reference star (2MASS J06303188+2942273) in the FOV. Nonetheless, the similar spectral type and K magnitude of the reference to WASP-12, and the lack of correlation between their fluxes and centroid positions on the detector (see \S\ref{wasp12}) still permit reliable flux calibration.  
Changing the exposure times during the observations can also introduce additional systematics to the data. However, since the flux of the target star is calibrated with the reference star, both stars have exactly the same exposure times, and the flat field and dark frames are well determined by averaging hundreds of images together,  seeing fluctuations are in principle the dominating systematics (although other systematics may also exist). Substantial amount of ``red noise" is thus relatively less likely to be introduced with different exposure times (see \S\ref{wasp12}). Therefore, the most significant effect of different exposure times is larger scatter of the data points with longer exposures due to more seeing fluctuation\footnote{This effect can be seen at the beginning of the two observations (see \S\ref{wasp12}).}.  

  For aperture photometry, we experimented with 48 aperture sizes with a step of 0.5 pixels. An aperture with radius of 19 pixels for 2010 November 26, and radius of 17 pixels for 2010 November 27 presents the smallest scatters in their out-of-eclipse and in-eclipse data, and thus is used as the final aperture in the photometry. We have also tested the eclipse depths with different aperture sizes for both nights in later analysis. Due to the good flat fielding of the data and no correlation of flux with centroid positions on the detector,  eclipse depths are stable and consistent for radii within [-3, +5] pixels of the best aperture for both nights.  A sky annulus with 29-pixel inner radius and 20-pixel width is used for background estimation.  Different annulus ranges and sizes have also been explored, and indicate consistent results. The timestamp (in UTC) of each image was first converted to BJD$_{UTC}$. Since \citet{Croll2011} has found no evidence of precession for WASP-12b, we thus calculated the orbital phases based on the non-precession ephemerides of \citet{Campo2011}, i.e., period=1.091424days, and transit epoch T$_0 (BJD_{UTC})$=2454508.97686.

\begin{figure}[t]
\begin{center}
 \includegraphics[width=3.2in, angle=90]{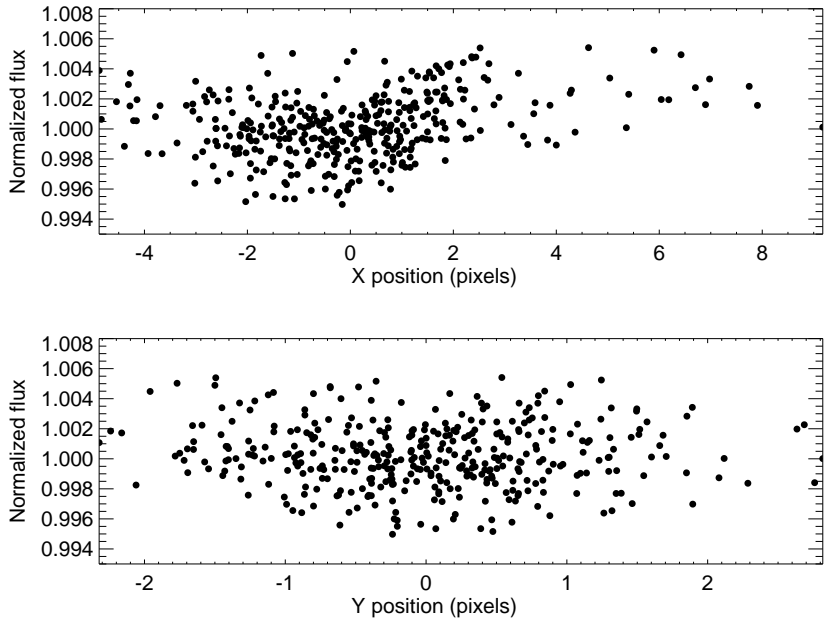}   
 \caption{Flux of CoRoT-1 as a function of X and Y positions of its centroid. The top panel shows the correlation of flux with X positions, while the bottom panel shows the correlation with Y positions. The linear correlation coefficient is 0.3 for the top panel and 0 for the bottom panel.}
\label{xy}
\end{center}
\end{figure}

\begin{figure}[t]
\begin{center}
 \includegraphics[width=2.6in, angle=90]{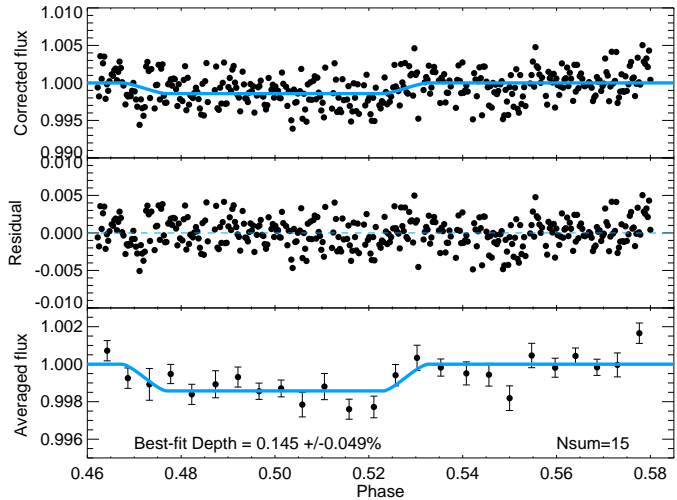}   
 \caption{Final corrected light curve of CoRoT-1b. The first panel shows the X-position de-correlated light curve of CoRoT-1b. The second panel shows the residual of the best-fit. The bottom panel shows the 15-points averaged light curve. The solid blue lines indicate the best-fit light curve of CoRoT-1b. Error bars of the points are calculated from the scatter of the data used for averaging. The 3-$\sigma$ percentile of the eclipse depth is between 0.033\% -- 0.235\%.}.
\label{lc}
\end{center}
\end{figure}

\begin{figure}[t]
\begin{center}
 \includegraphics[width=3.2in, angle=90]{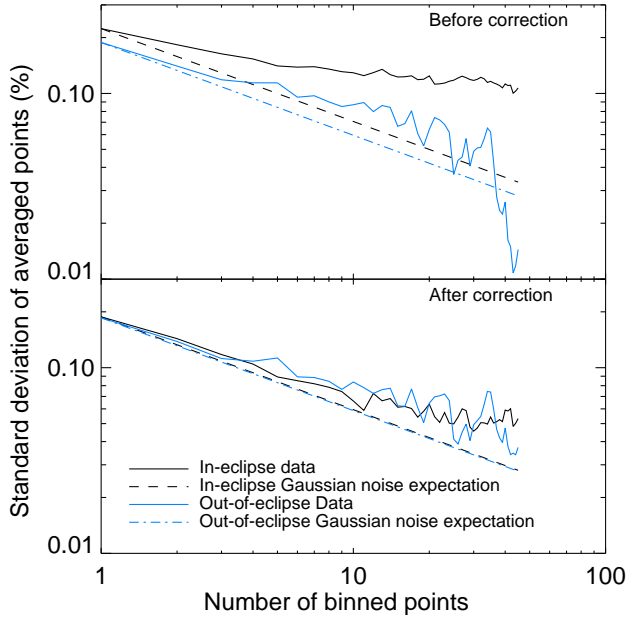}   
 \caption{Comparison of CoRoT-1b's noise level with Gaussian expectation. The top panel shows the standard deviation of the data as a function of binned points before the X-position de-correlation, while the bottom shows the standard deviation after the de-correlation. The solid lines indicate the noise levels of the actual data, while the dashed and dot-dashed lines indicate the Gaussian noise expectation.}.
\label{corot1_bin}
\end{center}
\end{figure}

\section{Analysis and results}
\label{result}

\subsection{CoRoT-1b}
\label{CoRoT}
After normalizing the time series of CoRoT-1 with the reference light curve, we still see large contaminations of correlated systematics (``red noise") in the data.  As can be seen in Figure \ref{flux}, the large-scale structures of the time series are highly correlated with the centroid positions of the star, possibly caused by a combination of  inter-pixel fluctuations due to imperfect flat-fielding and other systematic factors,  which cannot be corrected by the reference light curve since the individual fluctuations of the reference stars are averaged out.  Figure \ref{xy} shows the correlations of CoRoT-1's flux with the X and Y positions of its centroid. A clear trend and correlation is visible for the X positions (top panel), although the correlation coefficient is low due to the low signal-to-noise of the data. The Y positions do not appear to have strong correlation with the flux.  

To correct for these systematic trends, we apply a simple de-correlation function of: $f = 1 + a_1 X$, together with a linear slope of the form: $1 + a_2 t$, where $f$ is the flux, X is the positions of the centroid, and $t$ is the orbital phase of each measurement. The term for Y positions is not included due to the fact it has little or no correlation with the flux \footnote{We have also investigated the correction term for Y with the Bayesian Information Criterion and confirmed that the de-correlation function does not prefer such a term for Y positions}. Because of the insufficient out-of-eclipse baselines, we fit all the in-eclipse and out-of-eclipse data together. To account for the eclipse signature, we also fit simultaneously a light curve with the systematic corrections. The light curve is generated following the prescription of \citet{Mandel2002}, assuming uniform bodies without limb-darkening. 
The stellar and planetary parameters  for the light curve (R$_p$, R$_{star}$, inclination, and semimajor axis) are adopted from \citet{Gillon2009}.  
The free parameters in the least-square fit are: the eclipse depth, the level of the out-of-eclipse baseline, and the coefficients $a_1 \& a_2$. The known durations of ingress and egress are maintained in the fit. The center-of-eclipse timing is also fixed to phase=0.50 due to the fact that the orbital eccentricity of CoRoT-1b  is consistent with zero \citep{Rogers2009, Deming2011}, and our data have insufficient pre-ingress baseline for a robust constraint of $e\cos\omega$. 

We employed the Levenberg-Marquardt (LM) algorithm \citep{Press1992} for the least-square fit. 
To ensure that we find the global minimum instead of local minima, we searched  the parameter space extensively with a fine grid of starting points on top of the least-square fit. The grid has  a few hundred steps for each parameter.   The fact that most starting values on the grid converge to the same minimum suggests that we indeed have found the global minimum. 
The data points are uniformly weighted such that the  $\chi^2_{\nu}$ is nearly 1.0.  The  global best-fit light curve gives an eclipse depth of 0.145\%$\pm$0.028\%. The best-fit model is shown by the solid line in the second panel of Figure \ref{flux}. The X-position de-correlated data, the residuals of the best-fit and the averaged data are shown in Figure  \ref{lc}.  Figure \ref{corot1_bin} compares the noise level of CoRoT-1b with the Gaussian noise expectation before and after the X de-correlation. The in-eclipse data before the correction (top panel) suggest excessive ``red noise" above the Gaussian expectation. However, the noise level  reduced significantly after the correction and becomes more consistent with the out-of-eclipse data that are not affected much by the centroid drift (see Figure \ref{flux}). The Gaussian noise level (black dashed line) of the in-eclipse data is also improved after the correction.  The out-of-eclipse noise level (blue solid line), although is only slightly  improved, becomes more stable than before at larger bin numbers. Overall, the de-correlation has indeed reduced the systematics significantly. The average photon noise limit of a single exposure is about 0.049\%. Our final precision is thus about 4 times of the photon noise limit.

To verify if the X-position de-correlation function is truly preferred, we apply the Bayesian Information Criterion\footnote{BIC has been widely used for model identification and selection. Reduces in $\chi^2$ or maximum likelihood are penalized for the number of free parameters in BIC. Thus, the model with lower BIC value is generally preferred. }  (BIC)\citep{Liddle2007, Croll2011} for models with and without the de-correlation function. The result indicates that the model with the de-correlation function has a lower BIC value of 341.3, while the model without de-correlation has a higher BIC value of 361.2, suggesting that the X-position de-correlation function is indeed superior and preferred.

We have also investigated the possibility of a quadratic term for the out-of-eclipse baselines using BIC. The results are dependent on the baselines and do not completely justify a quadratic term\footnote{We have conducted two BIC tests, one with the post-egress baseline only, and the other with both the pre-ingress and post-egress baselines. The first test definitely prefers a linear model for the post-egress baseline (i.e., significantly lower BIC value). The second test marginally prefers a quadratic baseline. However, the quadratic fit is strongly leveraged by a few points (6) at the very end of the observation that have large scatters due to deteriorated seeing. If these points are excluded, the BIC once again strongly prefers a linear baseline. Therefore, a quadratic baseline cannot be completely justified. (When including these points, the quadratic baseline results in a best-fit eclipse depth of 0.102$\pm$0.033\%, still consistent with our final result.) }. In addition, the pre-ingress baseline is also too short to allow a reliable quadratic model. Thus,  we prefer a linear baseline model instead to avoid possible erroneous corrections an inaccurate quadratic term may introduce. A linear baseline can be more reliably determined since the in-eclipse data have also been used in the joint fit. We also note here that a quadratic term, if exists, should only affect the data slightly in large time scale; and the time correlated systematics caused by the quadratic term are taken into account in the following error analysis.


To examine the statistical significance and robustness of the eclipse depth and to estimate its error, we conduct 2 statistical tests. We first apply the standard bootstrapping technique \citep{Press1992}. In each bootstrapping iteration, we uniformly resample the data with replacement. Typically, a $\sim37\%$ of the original data points are randomly duplicated in each sample. For each new sample, we re-fit the X de-correlation function and the linear slope simultaneously with the light curve  to determine the eclipse depth, using the aforementioned grid search and LM minimization. This technique is suitable for unknown distributions like our case, and can robustly test the de-correlation model and the distribution of the parameters. A total number of 1500 iterations are performed and the resulting distribution of the eclipse depth is nearly Gaussian, with a median and 1-$\sigma$ deviation of 0.146\%$\pm$0.027\%, highly consistent with the previous best-fit. 

For the second test, we use the ``prayer-bead" residual permutation method \citep[and references therein]{Winn2008}. In brief, we subtract the best-fit model from the data and shift the residuals pixel-by-pixel. The shifted  residuals are then added back to the best-fit model to simulate a new set of 
data. The same de-correlation function and light curve are then employed to re-fit the new data for each iteration. We also reverse the residuals and iterate this process again,   resulting in a total number of 751 iterations (i.e., 2N-1, where N=376 is the number of data points). 
This method maintains the time-correlated errors and is therefore another robust way of testing our fit. Due to the un-corrected ``red noise" in the residual, the eclipse depth shows larger scatter in this test, and the final distribution is top-flat. The resulting median depth and 1-$\sigma$ error 
is 0.148\%$\pm$0.049\%, and the 3-$\sigma$ percentile is from 0.033\% to 0.235\%, suggesting the eclipse depth is detected at 3-$\sigma$ significance and is consistent with previous results. The results of these analysis are summarized in Table \ref{tab1}.

As a final test and cross check of our eclipse signal, we conduct a least-square fit to the original data without the X-position decorrelation but only a light curve and a linear slope. The resultant best-fit depth is 0.125\%$\pm0.029\%$, 
while the result from bootstrapping is 0.127\%$\pm0.028\%$, both are consistent with the previous results within error bars. The residual permutation method is also applied in this test, leading to a depth of 0.121\% with a 3-$\sigma$ percentile between 0.019\% - 0.029\% -- still consistent with the other tests.
The larger error in this permutation test is expected because of the excessive un-corrected ``red noise". 

Based on the above tests and their consistency, we conclude that our detection of the eclipse of CoRoT-1b is real and has at least 3-$\sigma$ significance. We report the final eclipse depth as 0.145\%$\pm$0.049\% based on the result from the original best-fit and the largest error from the residual permutation test (see Table \ref{tab1}).

\subsection{WASP-12b}
\label{wasp12}
\begin{figure}[t]
\begin{center}
 \includegraphics[width=2.6in, angle=90]{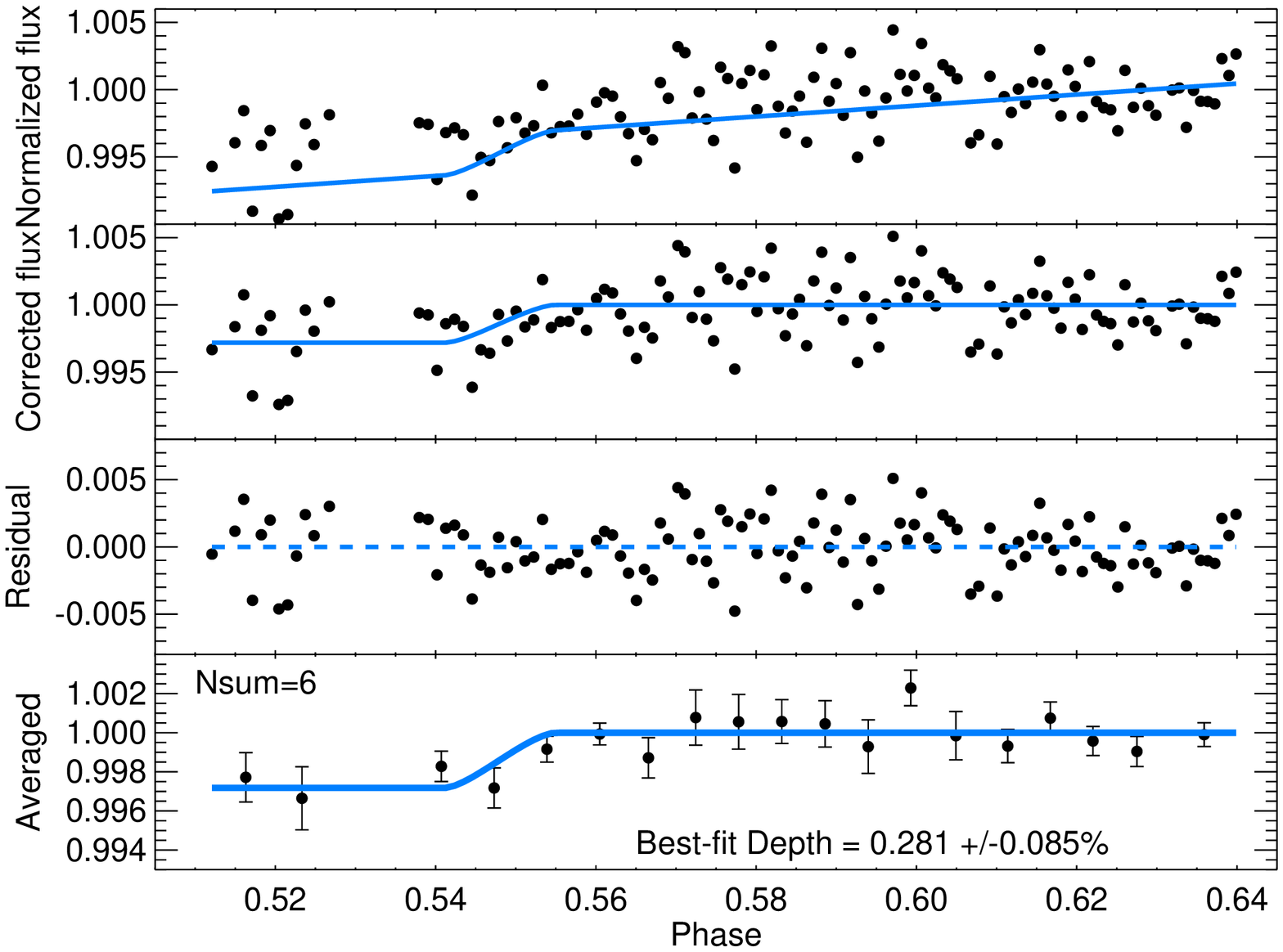}   
  \includegraphics[width=2.6in, angle=90]{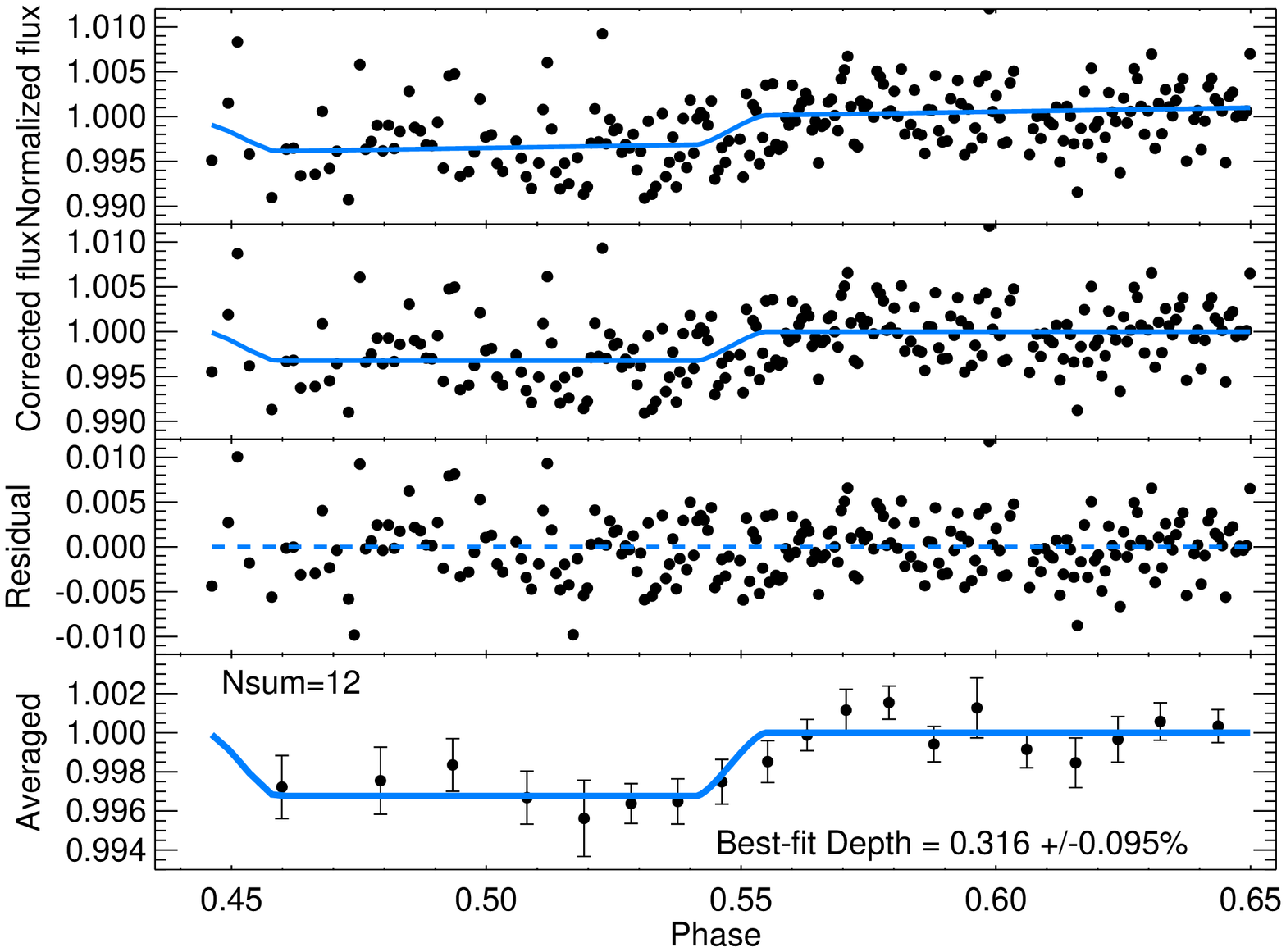}   
 \caption{Light curves of WASP-12b. The top panels show the flux of WASP-12b normalized to the reference star, overplotted with the best-fit background correction models and light curves. The middle panels show the background-corrected flux with the best-fit light curves. Residuals of the best-fit are shown in the third panels. The averaged data are shown in the bottom. Error bars are calculated from the scatter of the data used for averaging. }.
\label{lc2}
\end{center}
\end{figure}

\begin{figure}[t]
\begin{center}
 \includegraphics[width=2.35in, angle=90]{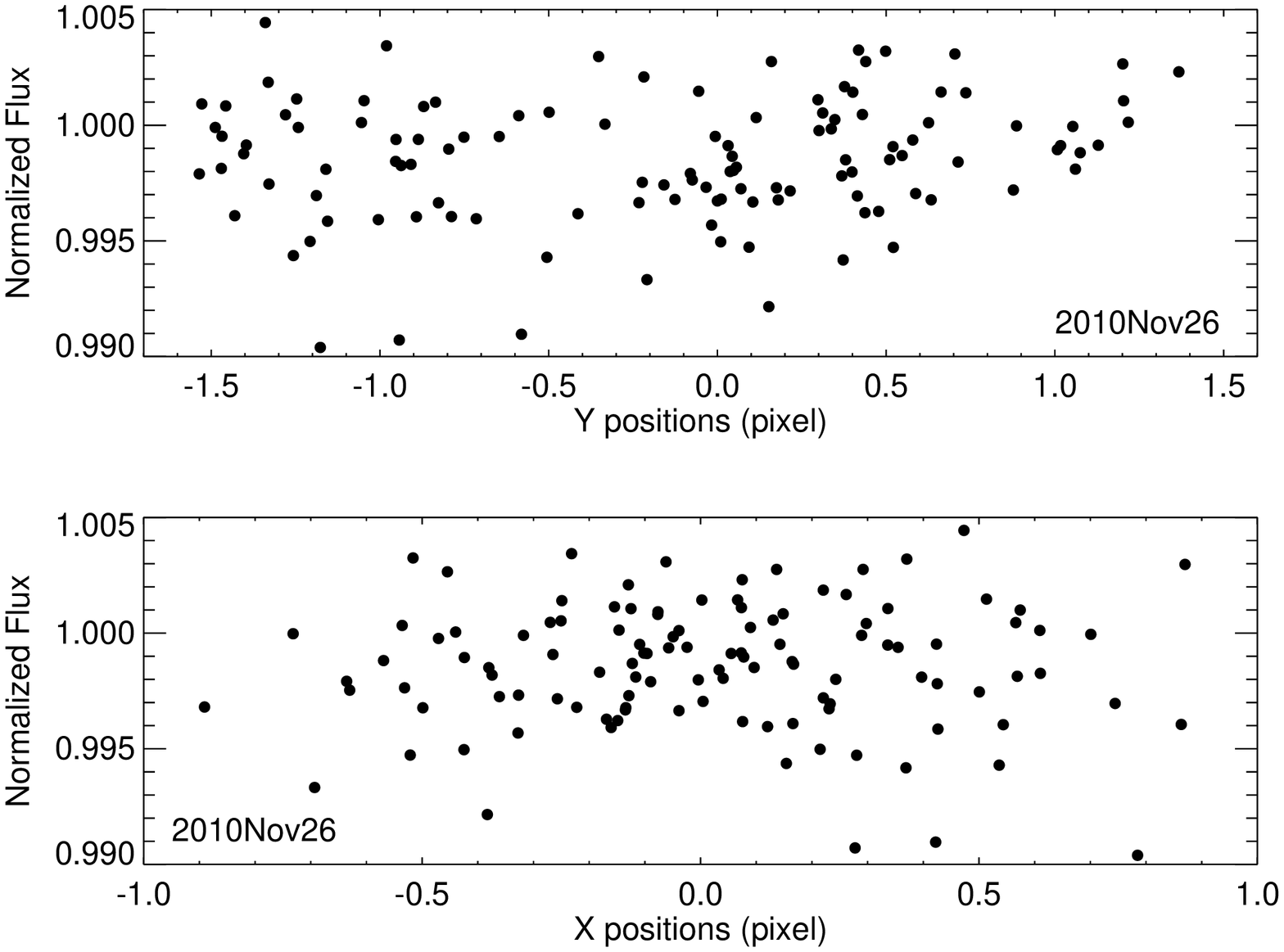}   
  \includegraphics[width=2.35in, angle=90]{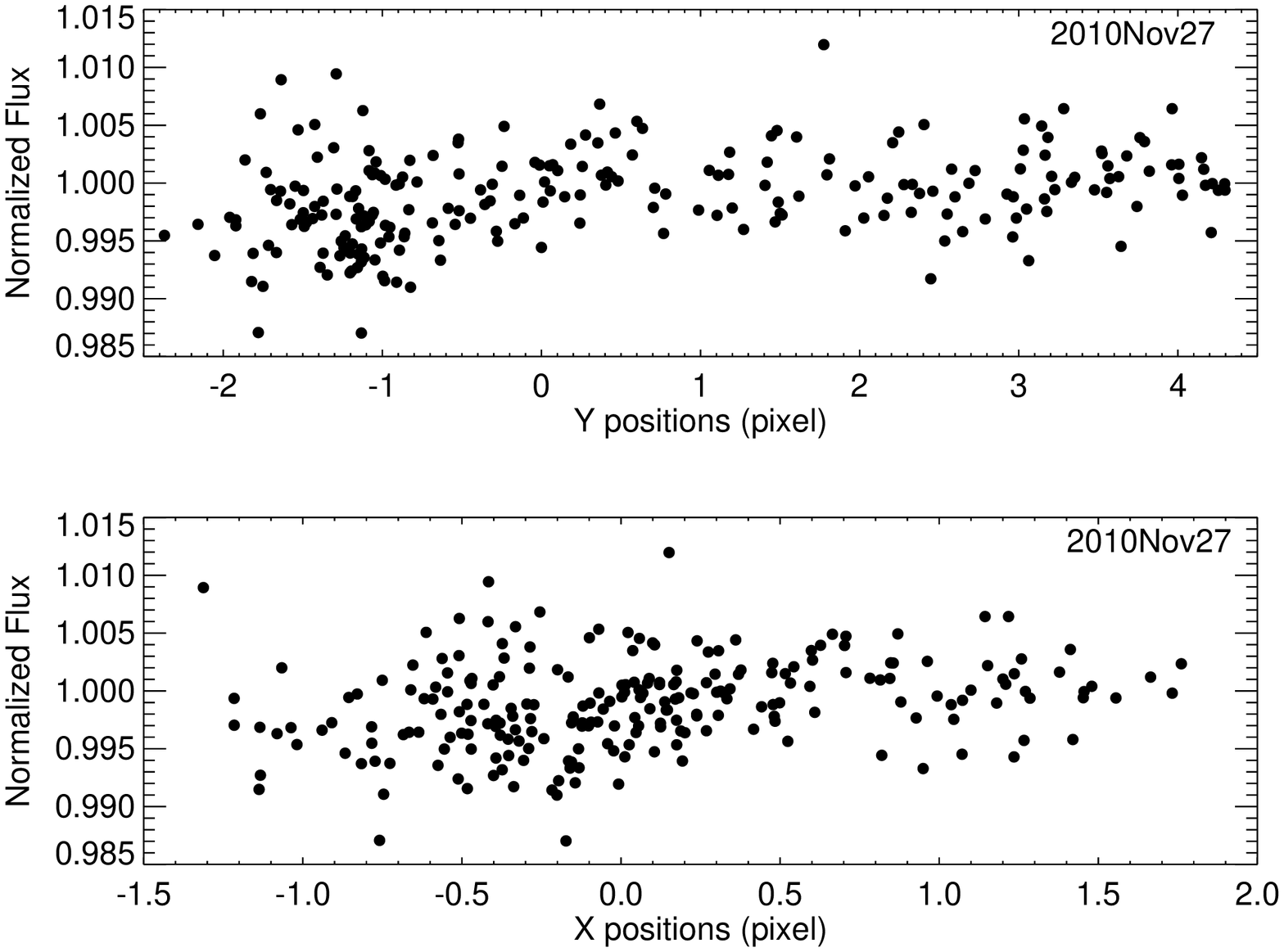}   
 \caption{Flux of WASP-12b as a function of X and Y positions of its centroid. The left two panels show the correlation for 2010 November 26, while the right panels show the correlation for 2010 November 27. Y positions (top panels) correspond to R.A. for MDM/TIFKAM, while X positions (bottom panels) correspond to Declination.   
}
\label{wasp12_xy}
\end{center}
\end{figure}

The two light curves of WASP-12b are analyzed in a similar way as CoRoT-1b. The reduced light curves of WASP-12b after normalizing with the reference star are shown in the top panels of  Figure \ref{lc2}. 
To correct for the systematics, we first investigate if centroid de-correlation functions are preferred by BIC. The calculations show that models without centroid de-correlation have lower BIC values than those with X or Y de-correlations for both nights, suggesting that simpler models are preferred. 
Figure \ref{wasp12_xy} shows the correlations of the reduced flux of WASP-12b (i.e., shown in top panels of Figure \ref{lc2}) with the X and Y positions of its centroid. The figures suggest there is no obvious correlations between its flux  and centroid positions. Therefore, no X or Y de-correlation terms are required in the models, as already suggested by the BIC tests.  

To investigate if quadratic terms are preferred in the out-of-eclipse baseline models of WASP-12b, we also calculate the BIC values for each night. Since there is no pre-ingress baseline for either night, only the post-egress baseline is used. It turns out that for both nights, linear background models are strongly preferred to quadratic models based on their lower BIC values. Therefore, in our final analysis of the WASP-12b data, we only apply a simple linear baseline model together with a light curve. 

Because WASP-12b's orbital eccentricity is consistent with zero \citep{Croll2011} and we have no pre-ingress baseline, we also keep the center-of-eclipse timing fixed to phase=0.5, and maintain the known durations of ingress and egress in the fit. Stellar parameters for the model light curves are adopted from \citet{Hebb2009}. An extensive grid of the starting parameters is also applied on top of the LM minimization to ensure we find the global minimum.  The data points are weighted uniformly such that $\chi_{\nu}^2$ is about 1. The final best-fit eclipse depth is 0.281\%$\pm$0.085\% for 2010 November 26, and  0.316\%$\pm$0.079\%  for 2010 November 27. 

\begin{figure}[t]
\begin{center}
 \includegraphics[width=2.6in, angle=90]{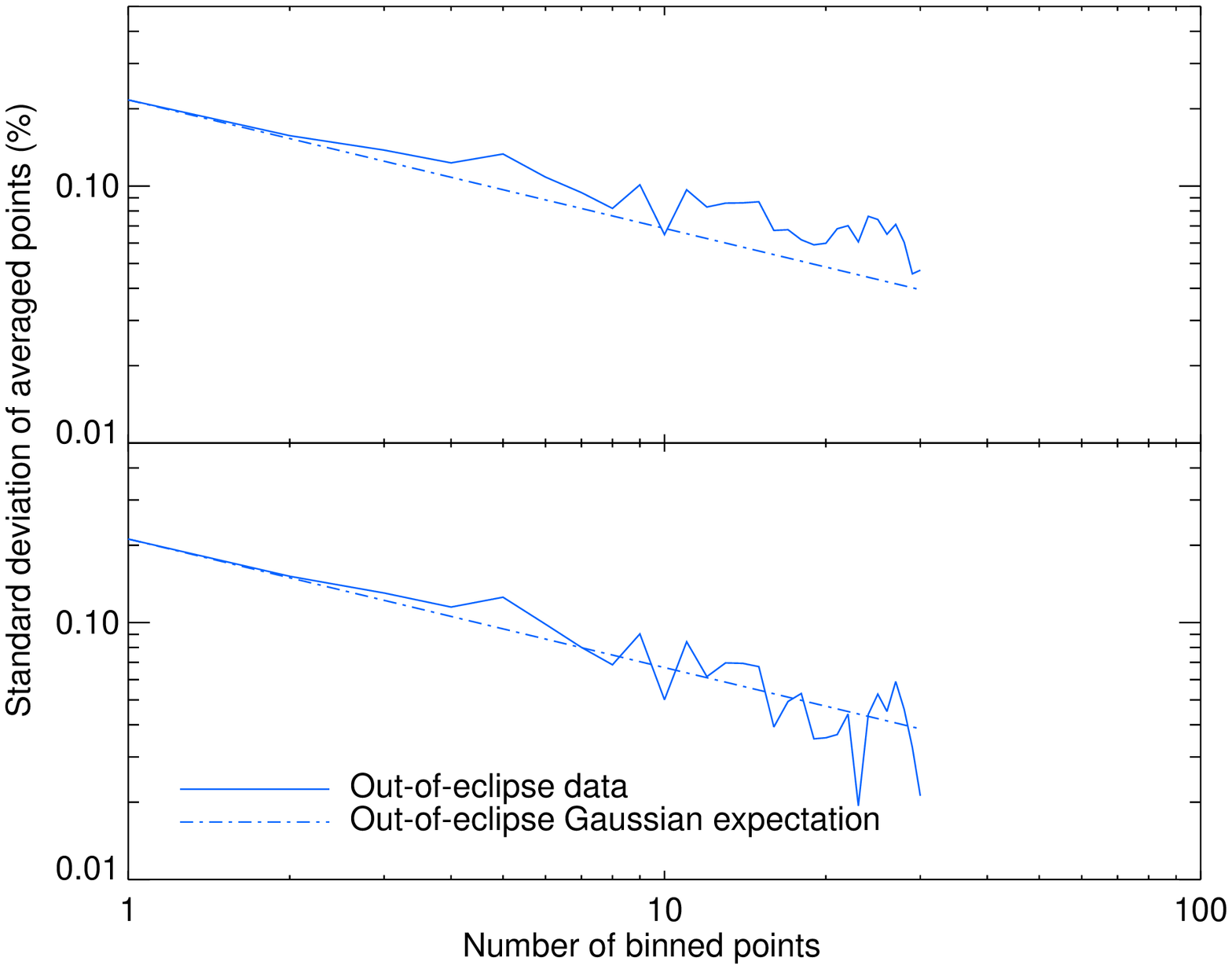}   
  \includegraphics[width=2.6in, angle=90]{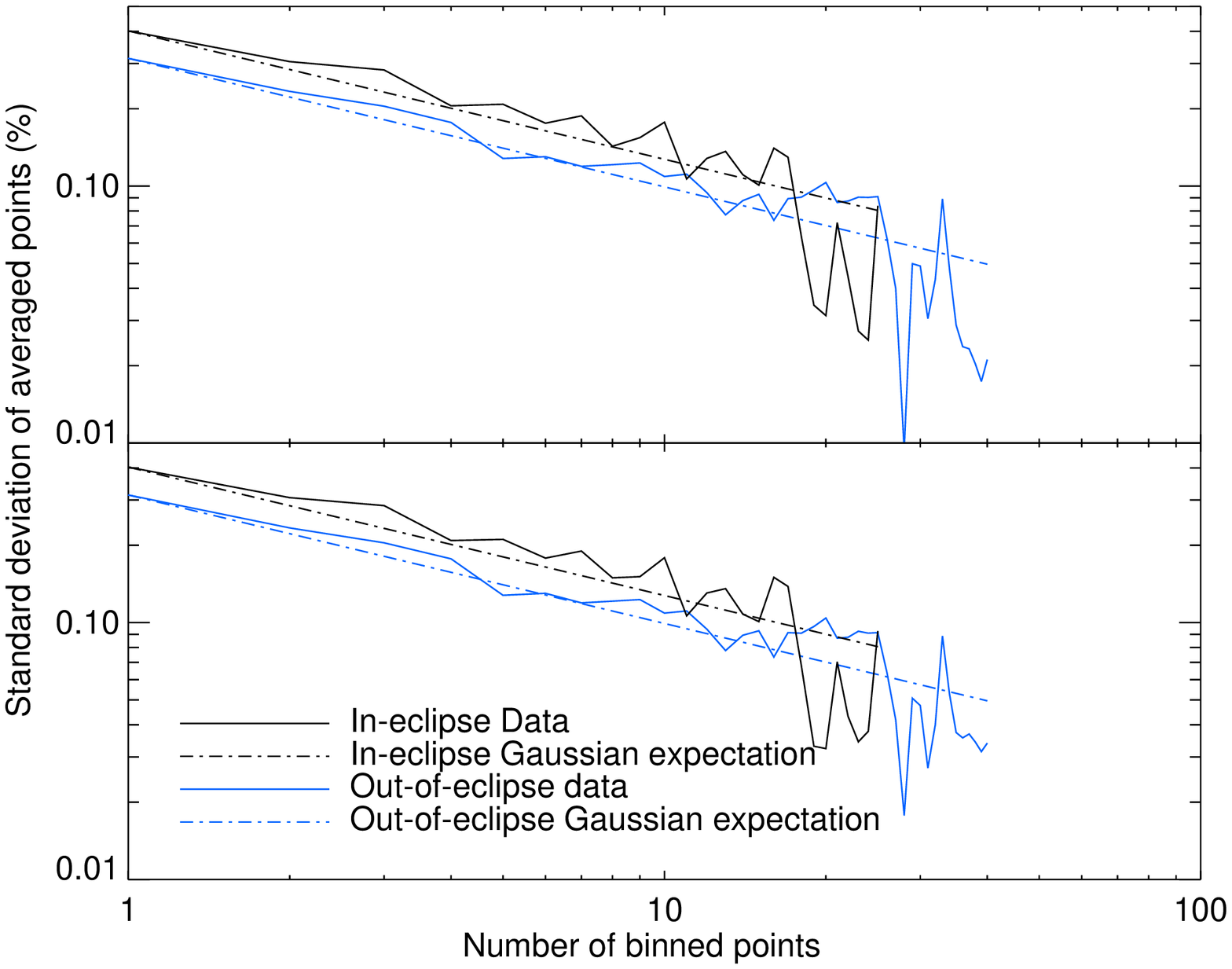}   
 \caption{Comparison of WASP-12b's noise level with Gaussian noise expectation. The top figure shows the standard deviations of the out-of-eclipse data as a function of binned points for 2010 November 26. The bottom figure shows the standard deviations of both the in-eclipse data and the out-of-eclipse data for 2010 November 27.
The solid lines indicate the binned down noise levels of the actual data, while the dot-dashed lines indicate the Gaussian noise expectation. The in-eclipse data of 2010 November 26 are not shown due to the small number of points. 
}
\label{wasp12_bin}
\end{center}
\end{figure}

The best-fit models, including the light curves, are shown in the top panels of Figure \ref{lc2} (blue lines). In the two top panels, both nights show a linear systematic trend in the background, which can be caused by large time scale seeing variation and/or drift of the  thermal background. 
The middle panels of Figure \ref{lc2} show the background-corrected data along with the best-fit light curves. Corresponding residuals and the averaged data are shown in the third and the bottom panels. Figure \ref{wasp12_bin} compares the noise level of WASP-12b with the Gaussian noise expectation before and after the background correction. Overall, there is very low systematic ``red noise" in both nights, and their standard deviations follow the Gaussian expectations closely. The bottom panel of 2010 November 26  shows that after the background correction, the noise level is effectively reduced to the Gaussian expectation. For the 2010 November 27  data, although the noise level stays nearly the same, the binned standard deviation becomes more stable at large bin numbers after the correction.

To examine the statistical significance and robustness of the model, we also implement the same bootstrapping and residual permutation tests for WASP-12b. A total number of 1500 iterations are carried out  for the bootstrapping test, resulting in a Gaussian-like distribution with a median depth and 1-$\sigma$ deviation of 0.290\%$\pm$0.085\% for 2010 November 26 and 0.329\%$\pm$0.077\% for 2010 November 27. The residual permutation tests  have  235 and 469  iterations for November 26 and November 27 respectively, resulting in a depth of 0.279\%$\pm$0.077\% for the former and 0.318\%$\pm$0.095\% for the latter. The results are summarized in Table \ref{tab1}.
The eclipse depths from both nights and all three methods are consistent with each other and all suggest better than 3-$\sigma$ significance. In addition, since the systematics of the two nights (e.g., airmass, seeing variation, thermal background, pointing drifts, observation timing and duration, etc.) are very different, and the only common feature of them is the eclipse signal, we conclude that the above detections are real.  Finally, we use the eclipse depths from the least-square fits and the largest error of the 3 methods to combine the two nights together. The final joint eclipse depth of WASP-12b is 0.299\%$\pm$0.064\% (4.7-$\sigma$), consistent with the $Ks$ band result of \citet{Croll2011}, 0.309\%$^{+0.013\% }_{-0.011\%}$.



\section{Discussion}
\label{discuss}

\begin{figure}[th]
\begin{center}
 \includegraphics[width=2.7in, angle=90]{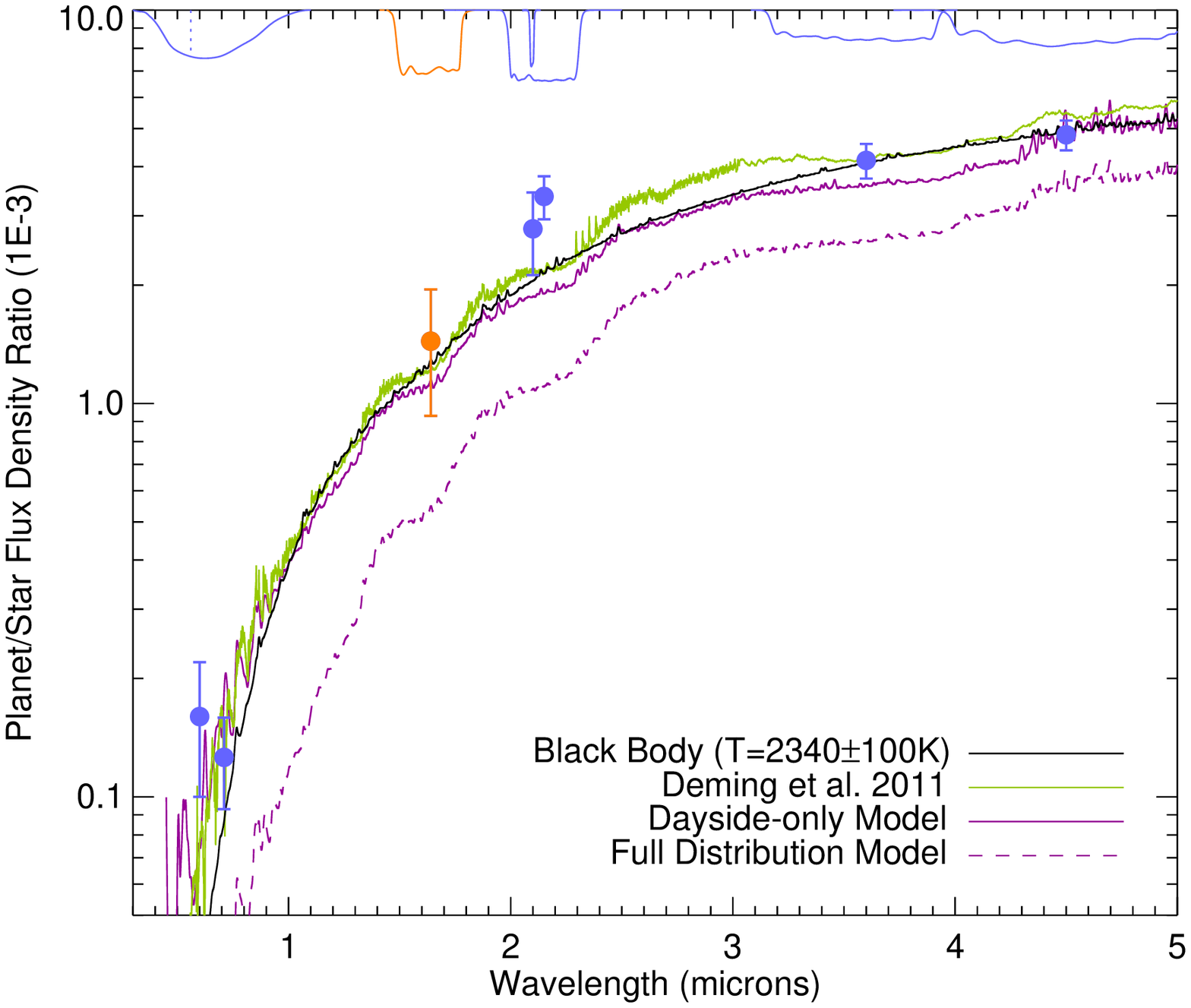}   
 \caption{Comparison of atmospheric models with CoRoT-1b data. The new $H$ band data is shown in orange. Previous measurements in other bands are shown in blue. The black line indicates the best-fit blackbody model. The temperature-inverted model from \citet{Deming2011} is shown by the green line.  The dayside only model from \citet{Gillon2009}  is shown by the solid purple line, while the model with full heat distribution over the day and night sides is  shown in dashed purple line. The inverted transmission profiles of the measurements are shown on the top of the figure. The dotted line that intersects the CoRoT bandpass  on the top left corner indicates the cut-off of the blue channel used in \citet{Snellen2009}.}
\label{spec}
\end{center}
\end{figure}

We summarize our new $H$ band measurement of CoRoT-1b in Figure \ref{spec} along with other previous results. 
Our $H$ band measurement corresponds to  a brightness temperature of 2280$^{+190}_{-230}$K for CoRoT-1b, slightly higher than its equilibrium temperature of 2180K, assuming a zero Bond Albedo and no heat redistribution from the dayside to the nightside. The  temperature-inverted atmospheric models of \citet{Gillon2009} and \citet{Deming2011} are fully consistent with our result.  
Our measurement supports the conclusion of \citet{Gillon2009} that a model with uniform redistribution of stellar flux across the entire planet surface is too cool to match the data (the dashed line). We also find a blackbody of T=2380$\pm$100K with no heat redistribution best fits the data at all wavelengths, implying an isothermal  region across much of the photosphere and very inefficient transport of heat to the night side. This is also in agreement with the conclusions of \citet{Deming2011}. The addition of the new $H$ band measurement is not sufficient to differentiate the best-fit models of previous studies at other wavelengths. Thus, in order to put more stringent constraints to current models,   measurements with better precision at these or new wavelengths are necessary.

\section{Conclusions}
\label{conclusion}

We have made a new detection of the very hot Jupiter CoRoT-1b's thermal emission at the $H$ band. The 3-$\sigma$ detection suggests a eclipse depth of 0.145\%$\pm$0.049\%. This result is consistent with the conclusions of  previous studies that the planet probably has a thermal inversion layer at high altitude, and has an isothermal region with inefficient heat transport across its dayside and nightside.  

We have also detected the thermal emission of  WASP-12b at the  $Ks$ band on two nights, at a 4.7-$\sigma$ joint eclipse depth of 0.299\%$\pm$0.065\%. This result independently confirms the previous detection of \citet{Croll2011} with a different telescope and instrument, suggesting the robustness of both measurements and also validating our data reduction and analysis method. 

Although more precise measurements are still required to better constrain models for CoRoT-1b and WASP-12b, our detections of  the two planets' secondary eclipses have brought the Palomar 200-in telescope and the MDM 2.4m telescope to the inventory of telescopes that have demonstrated the capability of  detecting hot Jupiter's thermal emission from the ground. Together with  previous results made with other telescopes, these detections suggest that ground-based observations are now mature and becoming a widely available tool to characterize the atmospheres of hot Jupiters. 


\acknowledgments
We thank the anonymous referee for valuable comments and suggestions for the paper. 
We thank John Tolbin, Zhaohuan Zhu, and the Palomar supporting
staff for their help with the observations. Part of the research described in this paper was carried out at the Jet Propulsion Lab (JPL)/California Institute of Technology (Caltech).
M.Z. is supported by the NASA Postdoctoral Program at JPL. 
T.B. acknowledges support from NASA Origins  grants to  Lowell Observatory 
and support from the NASA High-End  Computing Program through the NASA Advanced Supercomputing  Division.
 S.H. is supported by NASA's Sagan Fellowship at California Institute of Technology.    
TIFKAM was funded by OSU and the MDM consortium, and NSF grant AST-9605012. 
The Palomar Hale Telescope is operated by Caltech, JPL, and the Cornell University.



Facilities: \facility{Palomar Hale 200in, MDM Hiltner 2.4m}.

\begin{deluxetable}{llcc}
\tabletypesize{\scriptsize}
\tablecaption{CoRoT-1b and WASP-12b eclipse depth and error estimate}
\tablehead{
\colhead{Target} & \colhead{Method} & \colhead{Eclipse Depth} & \colhead{3-$\sigma$ percentile} 
} 

\startdata

						& Least-square fit & 0.145\%$\pm$0.028\%   &  0.061\% -- 0.229\%  \\
CoRoT	-1b				& Bootstrap          &  0.146\%$\pm$0.027\%   &  0.065\% -- 0.227\%  \\
						& Residual Permutation & 0.148\%$\pm$0.049\% & 0.033\% -- 0.235\% \\
		& Final result        &  \multicolumn{2}{c}{0.145\%$\pm$ 0.049\%} \\
\tableline
\tableline

						& Least-square fit 		& 0.281\%$\pm$0.085\%	&  0.025\% -- 0.535\%  \\	
WASP-12b				&Bootstrap          		&  0.290\%$\pm$0.085\%	&  0.035\% -- 0.545\%  \\	
(2010 November 26) & Residual Permutation 	& 0.279\%$\pm$0.077\% 	& 0.124\% -- 0.422\% \\	
		&Final result        		&  \multicolumn{2}{c}{0.281\%$\pm$ 0.085\%}    \\   
 \tableline
		 & Least-square fit 		& 0.316\%$\pm$0.079\% 	& 0.079\% -- 0.533\% 	\\
WASP-12b		 &Bootstrap          		& 0.329\%$\pm$0.077\%	& 0.098\% -- 0.560\%	\\
(2010 November 27) &Residual Permutation 	& 0.318\%$\pm$0.095\%	& 0.113\% -- 0.514\% 	\\
		&Final result        		&  \multicolumn{2}{c}{0.316\%$\pm$ 0.095\%} \\
\tableline
 WASP-12b&Joint solution			& \multicolumn{2}{c}{0.299\%$\pm$ 0.064\%} 
 \enddata
\label{tab1}
\end{deluxetable}




\end{document}